\documentclass[fleqn,twoside]{article}
\usepackage{amsmath}
\usepackage{espcrc2}
\usepackage{graphicx}

\newcommand{\fslash}[1]{\mbox{$\!\not\!#1$}}

\newcommand{\bold}[1]{\mbox{\boldmath ${#1}$}}
\newcommand{\agt}{\raisebox{.6ex}{$>$} \hspace{-.8em} \raisebox{-.6ex}
{$\sim$}}

\title{The phases of isospin asymmetric matter in the two flavor NJL model}

\author{S. Lawley,\address[cssm]{Special Research Centre
        for the Subatomic Structure of Matter, \\
        University of Adelaide, Adelaide SA 5005, Australia}
\address[jlab]{Jefferson Lab, 12000 Jefferson Avenue,
        Newport News, \\ VA 23606, U.S.A.}\thanks{Correspondence to: S.Lawley, E-mail:slawley@jlab.org}
        W. Bentz\address{Department of Physics, School of Science,
        Tokai University \\
        Hiratsuka-shi, Kanagawa 259-1292, Japan}
	and A. W. Thomas\addressmark[jlab] }

\date{ }

\begin{document}

\begin{abstract}
We investigate the phase diagram of isospin asymmetric matter at T=0 in
the
two flavor Nambu-Jona-Lasinio model. Our approach describes the single
nucleon
as a confined quark-diquark state, the saturation
properties of nuclear matter at normal densities,
and the phase transition
to normal or color superconducting quark matter at higher densities.
The resulting equation of state of charge neutral matter and the structure
of compact stars are discussed.

\vspace{1pc}

\noindent
{\footnotesize PACS numbers: 12.39.Fe; 12.39.Ki; 21.65.+f; 97.60.Jd }

\end{abstract}

\maketitle

\section{Introduction}

\setcounter{equation}{0}

The Nambu-Jona-Lasinio (NJL) model \cite{NJL} is an effective theory
of QCD at intermediate energies where
pointlike interactions between quarks replace the full gluon mediated
description of quark interactions.
The model has been widely used to study cold dense quark matter
(QM) \cite{QM}, because in this region of the QCD phase diagram the
explicit gluonic effects are
expected to be minor. Furthermore, recent developments of the NJL model
have shown the possibility of a realistic description of single nucleons
and stable nuclear matter (NM) \cite{BT}. Hadronization techniques
\cite{BHIT} in principle enable us to study both NM and QM
within the framework of a single model, which also accounts for the
internal
quark structure of the free nucleon \cite{FAD}. The purpose of this
letter is to investigate the phase diagram of this model, extending
the work of ref. \cite{BHIT} to the case of isospin asymmetric matter,
and to present results for the equation of state (EOS) 
of charge neutral matter and the structure of compact stars \cite{GL}.

Many recent theoretical studies have suggested that matter at high
densities and low temperatures is in the color superconducting
quark matter (SQM) phase \cite{SQM}. The number of flavors present will
depend on the effective quark masses in QM. In particular the
behavior of the strange quark mass at high
density, which is not well known, plays a critical role in
determining the structure of the favored QM phase \cite{AL}.
In NJL-type models, the strange quark turns out to be heavy enough
to favor a transition to the 2-flavor color SQM state \cite{QM},
whereas the 3-flavor state occurs at still higher densities.
There are many recent investigations on the
dependence on the strange quark mass \cite{REF1A,REF1B,REF2} 
which tend to support 
this scenario, provided that the quark pairing interaction is strong enough. 

In the present work we also examine the effects of color superconductivity
on the phase diagrams and the EOS at high densities.
It is a first attempt toward the goal of describing possible mixed NM/SQM 
phases in one framework.
We will use a description which, in normal NM, avoids unphysical
thresholds
for the decay of the nucleon into quarks, thereby simulating the effect of
confinement \cite{BT}. 
We will show the resulting EOS including mixed phases, and discuss the
consequences for compact stars.

We derive both NM and QM phases from the flavor SU(2) NJL-Lagrangian,
\begin{align}
{\cal L} &= \bar{\psi}(i \fslash{\partial}-m)\psi +
G_{\pi} \left((\bar{\psi}\psi)^2 - (\bar{\psi}\gamma_5
{\bold \tau}\psi)^2\right)  \nonumber \\
&- G_{\omega} (\bar{\psi}\gamma^{\mu}\psi)^2
- G_{\rho} (\bar{\psi}\gamma^{\mu}{\bold \tau}\psi)^2 \nonumber \\
&+ G_s (\bar{\psi}\gamma_5 C \tau_2 \beta^A \bar{\psi}^T)
(\psi^T C^{-1} \gamma_5 \tau_2 \beta^A \psi) \,,
\label{lag}
\end{align}
where we show only the interaction terms relevant for our discussions
\footnote[1]{We note that every 4-fermi interaction Lagrangian can be
decomposed,
as in Eq.(\ref{lag}), into various $q{\bar q}$ and $qq$ channels by using
Fierz transformations \cite{FAD}. In the last term of Eq.(\ref{lag}),
$C=i\gamma_2 \gamma_0$ and
$\beta^A = \sqrt{3/2} \lambda^A$ ($A=2,5,7$) are the color $\bar{3}$
matrices.}.
Here $m$ is the current quark mass, $\psi$ is the flavor SU(2) quark
field,
and the coupling constants $G_{\pi}$,
$G_{\omega}$ and $G_{\rho}$ characterize the $q{\bar q}$ interactions in
the
scalar, pseudoscalar and vector meson channels, while $G_s$ refers to the
interaction in the scalar diquark channel.

The model is further specified by a regularization scheme, for which
we use the proper-time scheme \cite{PT} in this work. It is characterized
by an infrared cut-off ($\Lambda_{\rm IR}$) in addition to the usual
ultraviolet one ($\Lambda_{\rm UV}$).
In the vacuum, the parameters of the model are determined as follows:
We fix $\Lambda_{\rm IR}=200$ MeV, and choose
$\Lambda_{\rm UV}$, $m$ and $G_{\pi}$ so as to reproduce
$f_{\pi}=93$ MeV, $m_{\pi}=140$ MeV, and constituent quark mass
$M_0=400$ MeV via the gap equation at zero density.
We will set $M=0$ in QM because the quark mass is already very small in
the region where the transition to QM occurs.  We have found that
including the effective quark mass in this phase does not change the
structure of the phase diagrams and has little effect on the EOS.

\section{Nuclear matter}
\setcounter{equation}{0}
The nucleon is constructed as a quark-diquark bound state \cite{FAD},
making use
of the scalar diquark interaction term in (\ref{lag})
and the Bethe-Salpeter equation to get
the scalar diquark mass, $M_s$. The interaction with the spectator
quark is described by the quark exchange (Faddeev) kernel, for which
we use a momentum-independent approximation in the finite
density calculations reported in this letter. (This corresponds to
the ``static approximation'' of the Faddeev kernel --
see Refs.\cite{STAT,BT} for details.) The coupling constant
$G_s$ is chosen to reproduce the free nucleon mass, $M_{N0}=940$ MeV.

In the mean field approximation, the NM phase is characterized by
composite neutrons and protons, moving in scalar and vector mean fields.
There is also a non-interacting sea of
electrons in chemical equilibrium with the nucleons
($\mu_e = \mu_n - \mu_p $, where $\mu$ denotes the chemical potential)
\footnote{At some density the components may change,
as negatively charged kaon or pion condensates \cite{REF3} as well as muons
replace the electrons and more
massive hadron species replace nucleons \cite{BAYM}, but here we
confine ourselves to the most simple picture.}.
The form of the effective potential in the mean field approximation
has been derived for symmetric NM in Ref.\cite{BHIT} starting from the
quark Lagrangian (\ref{lag}) and using the hadronization method.
This can be easily extended to the isospin asymmetric case. It has the
form
\begin{eqnarray}
V^{\rm (NM)} = V_{\rm vac} +  V_{N} - \frac{\omega_0^2}{4 G_{\omega}}
- \frac{\rho_0^2}{4 G_{\rho}} - \frac{\mu_e^4}{12 \pi^2}\,, \label{vnm}
\end{eqnarray}
where
\begin{multline}
V_{\rm vac} = 12i \int \frac{{\rm d}^4 k}{(2\pi)^4}\,{\rm ln} \,
\frac{k^2-M^2}{k^2-M_0^2} \\
+ \frac{(M-m)^2}{4 G_{\pi}}
-  \frac{(M_0-m)^2}{4 G_{\pi}} \label{vac}
\end{multline}
is the vacuum term, and
\begin{multline}
V_N = -2 \sum_{\alpha={\rm p,n}} \int \frac{{\rm d}^3 k}{(2\pi)^3}\,
  \\
\hspace{15mm}\Theta(\mu^*_{\alpha} - E_N(k))\left(\mu^*_{\alpha} -
E_N(k)\right)
\label{vn}
\end{multline}
describes the Fermi motion of nucleons moving in the scalar and vector
mean fields. We used $E_N(k)=\sqrt{M_N^2 + k^2}$, where $M_N(M)$ is the
nucleon mass in-medium, which is obtained from the pole
of the quark-diquark T-matrix.  The effective chemical potentials
are defined as $\mu^*_{\alpha}=\mu_{\alpha}-3 \omega_0 \mp 3 \rho_0$
in terms of the mean vector fields
$\omega^0 = 2 G_{\omega} \langle {\rm NM}| \psi^{\dagger} \psi
|{\rm NM} \rangle$ and $\rho^0 = 2 G_{\rho} \langle {\rm NM}|
\psi^{\dagger} \tau_3 \psi |{\rm NM} \rangle$.

The constituent quark
mass, $M$, and the mean vector fields in NM are determined
by minimizing $V^{\rm (NM)}$ for fixed chemical potentials
$\mu_p$ and $\mu_n$.
The parameter $G_{\omega}$ is fixed by the requirement that the
binding energy per nucleon of symmetric NM
passes through the empirical saturation point (baryon density
$\rho_0 \equiv 0.16$ fm$^{-3}$ and $E_B/A=15$ MeV)
\footnote{We recall from Ref.\cite{BT} that $G_{\omega}$ is the only
free parameter for NM.  With the limited number of parameters
in this simple model it is not possible to ensure that the calculated
binding energy curve also has a {\em minimum} at the empirical
saturation point. Instead it occurs at $\rho=0.22$ fm$^{-3}$ and
$E_B/A=17$ MeV.}
, and the parameter
$G_{\rho}$ is adjusted to the empirical
symmetry energy ($a_4=32$ MeV at $\rho_0$).
The resulting values of the parameters are shown in the column
``NM'' of Table 1.

As compared to chiral models for point-nucleons, the important property 
which leads to saturation of the NM binding energy
in this approach is the positive curvature of the function $M_N(M)$, which
reflects the internal quark structure of the nucleon and which
works efficiently only if there are no unphysical thresholds for the
decay of the nucleon into quarks\cite{BT}. In our method, these thresholds
are avoided by the choice of the proper-time regularization scheme
with $\Lambda_{\rm IR}>0$ \cite{PT}.

\begin{table}[hbt]
\begin{center}
\begin{tabular}{|c||c|c|}
\hline
                        &  NM         & QM \\  \hline
m [MeV]                 &   16.93          & 17.08  \\
$G_{\pi}$[GeV$^{-2}]$   &  19.60           & 19.76 \\
$\Lambda_{\rm UV}$[MeV] &  638.5           & 636.7 \\
$\Lambda_{\rm IR}$[MeV] &  200.0           & 0 \\
$r_{\omega}\equiv G_{\omega}/G_{\pi}$  &  0.37     & 0 \\
$r_{\rho}\equiv G_{\rho}/G_{\pi}$     &  0.092   & 0 \\
$r_s\equiv G_s/G_{\pi}$    &  0.51     & free parameter \\  \hline
\end{tabular}
\end{center}
\caption{Parameters used for nuclear matter (left column) and
for quark matter (right column). The proper time regularization scheme
is used in both cases.  Because $\Lambda_{\rm IR}$ is set to zero in the
QM case, the parameters m, $G_{\pi}$ and $\Lambda_{\rm UV}$ differ
slightly from the NM values in order to obtain $f_{\pi}=93$ MeV,
$m_{\pi}=140$ MeV, and $M_0=400$ MeV.}
\end{table}

\section{Quark matter}
\setcounter{equation}{0}

The effective potential for QM in the mean field approximation,
allowing for the possibility of
diquark condensation, has the form
\begin{eqnarray}
V^{\rm (QM)} = V_{\rm vac} +  V_{Q} + V_{\Delta} -
\frac{\mu_e^4}{12 \pi^2} \label{vqm}
\end{eqnarray}
where the vacuum part, $V_{\rm vac}$, is given by (\ref{vac}) and
\begin{multline}
V_Q = -6 \sum_{\alpha={\rm u,d}} \int \frac{{\rm d}^3 k}{(2\pi)^3}\,
 \\
 \Theta(\mu_{\alpha} - E_Q(k))\left(\mu_{\alpha} - E_Q(k)\right)
\label{vq}
\end{multline}
describes the Fermi motion of quarks with chemical potentials
$\mu_u$ and $\mu_d$, and $E_Q(k)=\sqrt{M^2+k^2}$. The term $V_{\Delta}$
describes the effect of the pairing gap and is given by
\begin{multline}
V_{\Delta} = 2i \int \frac{{\rm d}^4 k}{(2\pi)^4}
\sum_{\alpha=+,-} \Big[ {\rm ln} \, \frac{k_0^2 -
(\epsilon_{\alpha}+\mu_I)^2}
{k_0^2 - (E_{\alpha}+\mu_I)^2} \\
+ {\rm ln} \, \frac{k_0^2 - (\epsilon_{\alpha}-\mu_I)^2}
{k_0^2 - (E_{\alpha}-\mu_I)^2} \Big] + \frac{\Delta^2}{6 G_s}\,,
\label{delta}
\end{multline}
where $\epsilon_{\pm}(k) = \sqrt{(E(k) \pm \mu_q)^2 + \Delta^2}$ and
$E_{\pm} = |E(k) \pm \mu_q|$. Here we used the isoscalar and isovector
combinations $\mu_q = (\mu_u + \mu_d)/2$
and $\mu_I = (\mu_u - \mu_d)/2$
\footnote{We mention that in principle one
needs a further chemical potential for color neutrality ($\mu_8$) in QM.
However, for the 2-flavor case $\mu_8$ turns out to be very small
\cite{QM,ABG}, and therefore we neglect it here for simplicity.}.

A detailed discussion of vector meson poles in Ref.\cite{BHIT}
has shown that the vector-type interactions should be set to zero in QM.
In addition, we set
$\Lambda_{\rm IR}=0$, as the infrared cut-off simulates
confinement effects which are not appropriate in QM.
  Also, as stated in the Introduction, we will set $M=0$ in QM, since the
effects of quark mass are small when constructing the phase diagrams and
the charge neutral EOS.
The resulting parameters are shown in the column ``QM''
of Table 1. In the main part of this paper we treat $G_s$ as a free
parameter in QM to investigate the dependence on the pairing strength, and
comment on the important question of consistency with the value derived
from the free nucleon mass at the end of Sect.4.

The gap $\Delta$ is determined by minimizing the effective potential
for fixed quark chemical potentials.
In the following discussions, we will distinguish the normal quark matter
(NQM) phase, which is characterized by $\Delta=0$, from the SQM phase
($\Delta>0$).

\section{Results}
\setcounter{equation}{0}

In order to construct the phase diagram, we compare the effective
potentials in the NM, NQM and SQM phases for several fixed chemical
potentials for baryon number and isospin ($\mu_B$ and $\mu_I$), which
are related to the chemical potentials of the particle species by
\begin{align}
\mu_{\alpha} &= \mu_B \pm \mu_I \,\,\,\,\,(\alpha={\rm p,n})\,,
\,\,\,\,\,\, \nonumber \\
\mu_{a} &= \frac{\mu_B}{3} \pm \mu_I \,\,\,\,\,(a={\rm u,d})\,,
\,\,\,\,\,\, \nonumber \\
\mu_e &= -2 \mu_I\,.
\end{align}

\begin{figure*}
  \begin{center}
    \includegraphics[width=5in,height=4.5in]{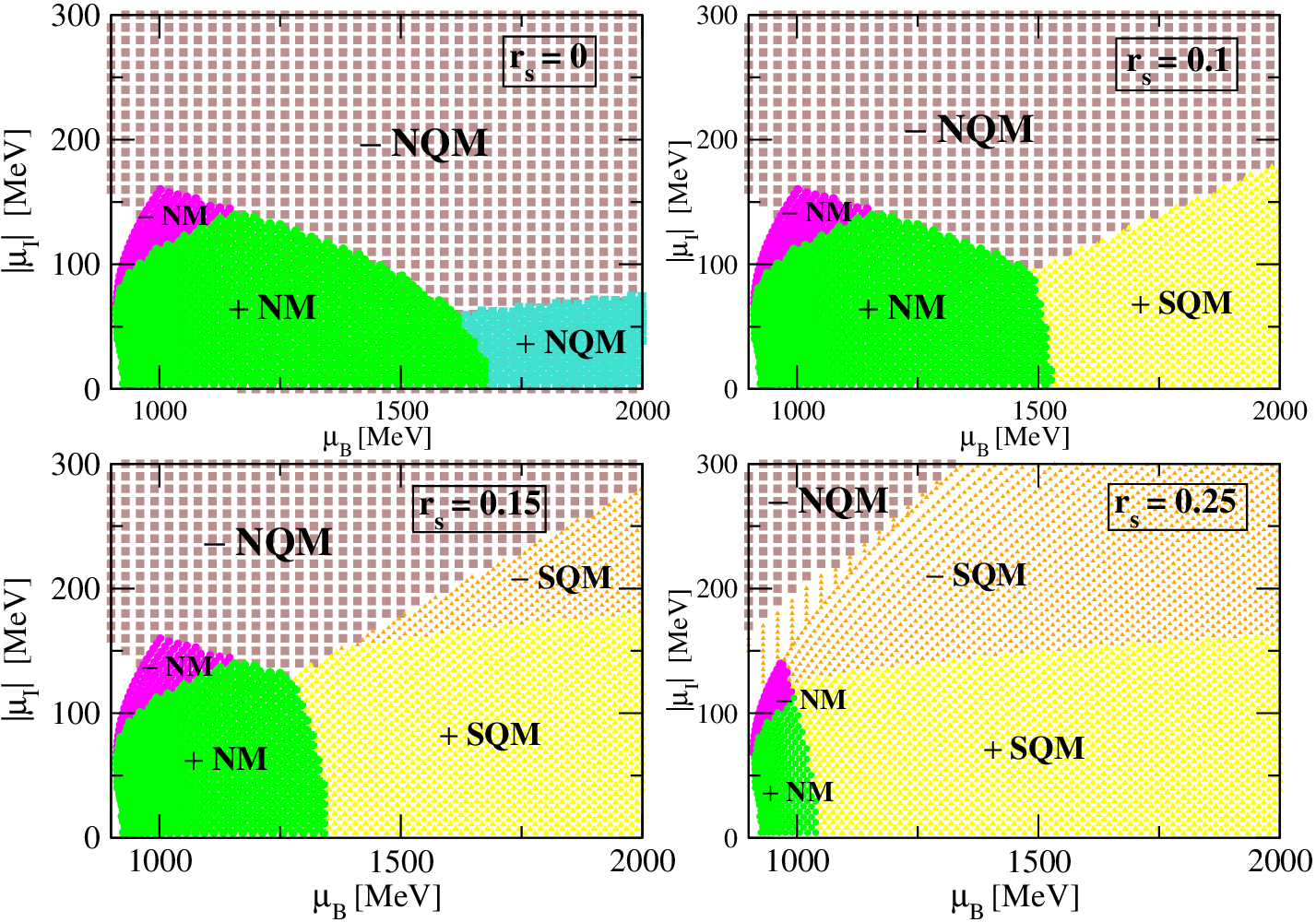}
    \caption{Phase diagrams in the plane of chemical potentials
$\mu_B,\,\mu_I$ for baryon number and
             isospin for various choices of $r_s$. The black, dark and
light regions correspond to
the NM, NQM and SQM phases, and the $\pm$ indicate the sign of the total
charge density including
            the electrons in chemical equilibrium.}
    \label{f1}
  \end{center}
\end{figure*}

\vspace{2mm}
\begin{figure*}
  \begin{center}
    \includegraphics[width=5in,height=4.5in]{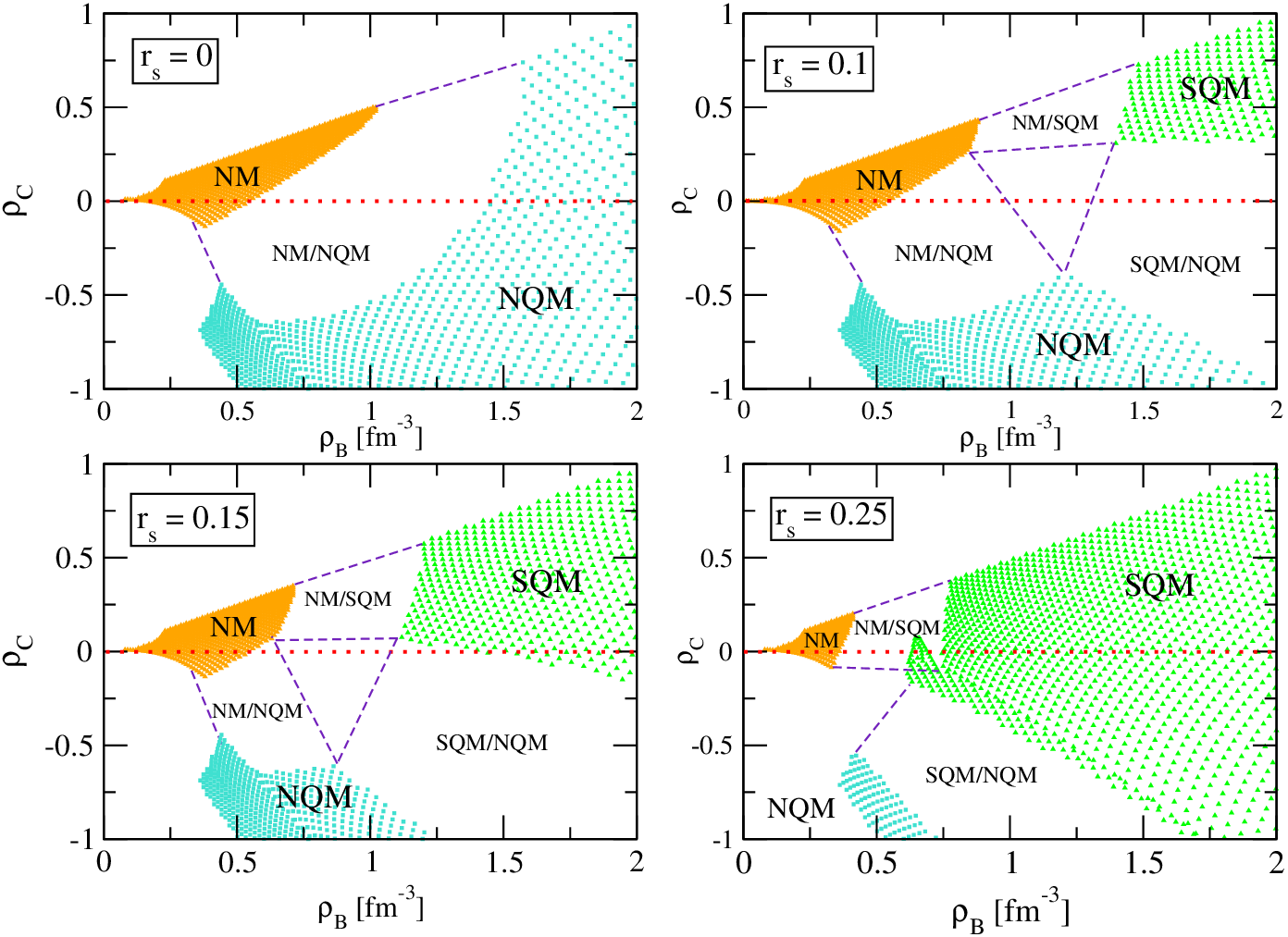}
    \caption{Phase diagrams in the plane of densities $\rho_B,\,\rho_C$
for baryon number and
             charge for various choices of $r_s$. The black, dark and
light regions correspond to
the NM, NQM and SQM phases, the white regions separated by the dashed
lines correspond to the
mixture of two phases, and the triangular region for the cases $r_s=0.1$
and $0.15$ involves a mixture
of three phases. The upper left regions in each diagram are left empty,
because we consider only
the case $\mu_I<0$. The dotted line indicates charge neutrality.}
    \label{f2}
  \end{center}
\end{figure*}

Fig.1 shows the phase diagrams for several choices of the pairing strength
in the SQM phase. The black, dark and light regions indicate the phases
with the
lowest effective potential, and the $\pm$ indicate
the sign of the total charge density\footnote{The density of
particle $A=p,n,u,d,e$ is obtained as
$\rho_A = - \partial V / \partial \mu_A$. In particular, the baryon and
charge densities in NM are
$\rho_B=\rho_p + \rho_n, \,\,\rho_c = \rho_p - \rho_e$, and in QM
$\rho_B = (\rho_u + \rho_d)/3, \,\, \rho_c = 2/3 \rho_u - 1/3 \rho_d -
\rho_e$.}.
The corresponding plots in the plane of baryon and charge density are
shown in
Fig.2. The mixed phases appear as white regions in this Figure. The phase
boundaries,
which are single lines in Fig.1, appear as two lines facing each other in
Fig.2.
By connecting the corresponding end points on the boundaries by straight
lines
(dashed lines in Fig.2), we can divide the white region into sections
which
correspond to mixtures of the two phases facing each other.
The $\mu_I = 0$ axes in Fig.1 correspond to the upper-most line running
from NM to QM
in Fig.2, and because we consider only the case $\mu_I<0$ the upper left
parts of the
diagrams in Fig.2 are left empty\footnote{In principle we could extend the
phase diagrams
to the region where $\mu_I < 0$ by admixing positrons instead of
electrons, but clearly the matter
in this part of the phase diagram is always positively charged and is not
relevant to the charge neutral EOS.}.

For each case in Fig.1, the charge
neutral EOS corresponds to the line separating the
positively and negatively charged regions.
For example in the first diagram ($r_{s} = 0$), the charge
neutral EOS begins in the pure NM phase, then there is a
mixed (+NM/-NQM) phase and finally a pure NQM phase.  Each point on
the boundary between the NM and NQM phases satisfies the Gibbs conditions,
since $P^{(NM)}(\mu_{B},\mu_{I}) = P^{(NQM)}(\mu_{B},\mu_{I})$. The EOS
in the mixed neutral phase is found by using the method of
Glendenning \cite{GL1}, that is, the volume fraction of the NM
phase is determined by the requirement of charge neutrality as
$x^{(NM)} = \rho_c^{(QM)}/(\rho_c^{(QM)}-\rho_c^{(NM)}),\,\,
x^{(QM)}=1-x^{(NM)}$. Physically this method implies that the mixed
phase begins with charge neutral NM, and then as the charge of NM becomes
increasingly positive, regions of negatively charged NQM form, such that
the mixed phase remains globally charge neutral\footnote{For such a
mixture one may calculate
what sizes and shapes are
favored for each component \cite{SHAPE}.}. The same sequence of
neutral phases can also be seen in the first diagram of Fig.2 by following
the
horizontal line $\rho_C=0$.

\begin{figure*}
\begin{center}
\includegraphics[width=5.5in,height=2.5in]{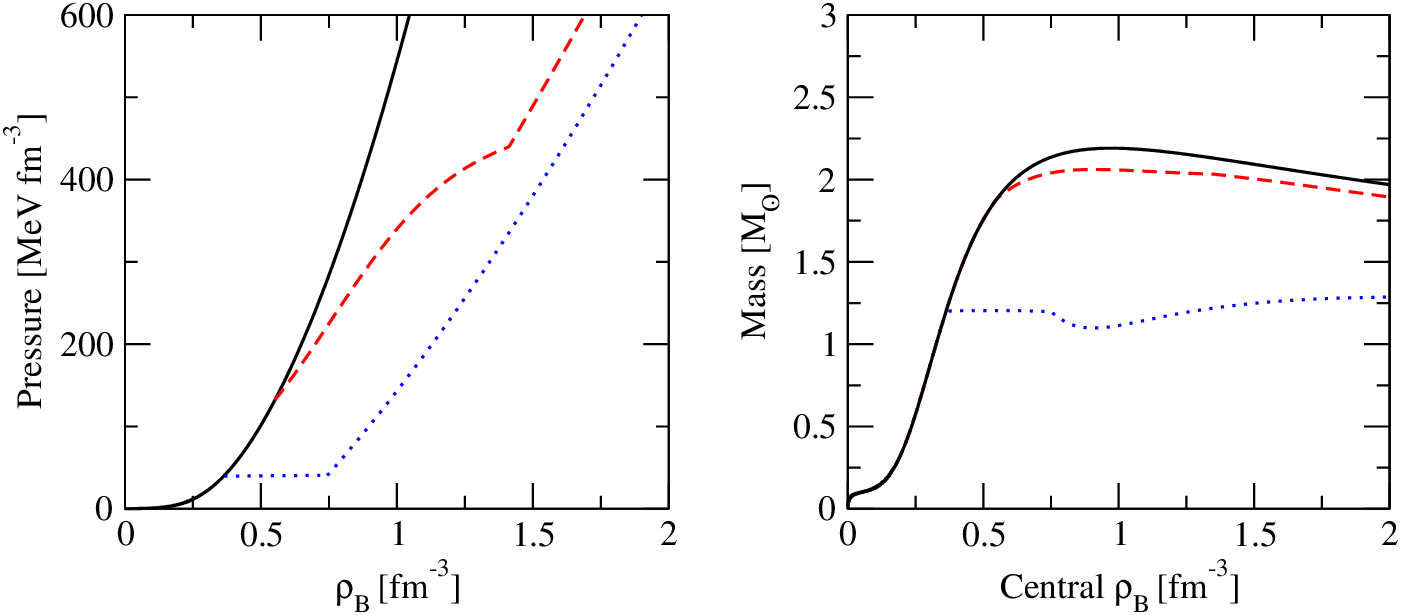}
\caption{Left figure: Charge neutral EOS (pressure against baryon density) 
for pure NM (solid line), the transition to NQM (case $r_s= 0$, dashed line),
and the transition to SQM (case $r_s= 0.25$, dotted line). Right figure:
The corresponding star masses as functions of central density.}
\label{f3}
\end{center}
\end{figure*}

As we increase the pairing strength in the QM phase, the
regions where SQM is the ground state extend. When $r_{s}=0.1$
we have a mixed (+NM/-NQM) phase, and then come to a triple point
where all three phases meet. This is the same situation as investigated
in Ref. \cite{TRIP},
and leads to a region of constant pressure in the EOS, where all three
phases are mixed.
This is illustrated in Fig.2 by the triangular region; that is, when the
$\rho_{c}=0$ line passes
through this region all three phases are present.
The volume fraction of SQM begins at zero on the left hand side of the
triangle and increases while the volume
fraction of NM decreases until it reaches zero on the right hand side of
the triangle.
The NQM phase occupies the remaining volume fraction, which varies
continuously between the boundaries of the triangle.
In this region the baryon density is increasing, while the pressure
remains constant.
However, within compact stars the pressure must always be decreasing as a
function of radial position.
 Thus the three component mixed phase cannot occupy any finite volume
within a star and there will be a
discontinuity in the star's density profile.

If we further increase the pairing strength to $r_s = 0.15$, a charge
neutral SQM state becomes possible.
This transition from NM to SQM involves three intermediate mixed phases
(NM $\to$ +NM/-NQM $\to$ +NM/-NQM/+SQM $\to$ -NQM/+SQM $\to$ SQM). Again,
the three component mixed phase occurs at the
triple point in Fig.1, which corresponds to the triangle in Fig.2.
At still larger pairing strengths the NQM phase becomes unfavorable and
the EOS involves just NM and SQM.
When $r_s=0.25$ we start with a neutral NM phase, and then enter a
(-NM/+SQM) mixed phase before arriving at the
neutral SQM phase. In Fig.1 the line along the mixed phase in this case is
only very short,
which means that the pressure changes in the mixed
phase are small. For ($r_s \agt 0.3$) the SQM
phase almost completely expels the NM phase,
which indicates an upper limit for the pairing strength which is
consistent with the findings
of Ref.\cite{BHIT} for the isospin symmetric case.

In Fig.3 we show two examples of the charge neutral EOS,
corresponding to the cases $r_s=0$ and $r_s=0.25$, and the resulting 
star sequences obtained by integrating the TOV equation\cite{TOV}. 
(The results
for the pure NM case are also shown for comparison.)
In the first case, the system goes to the NM/NQM mixed phase at
a baryon density around $3.4 \rho_0$  and the pressure of the
mixed phase increases with increasing density. For central densities
between $3.4 \rho_0$ and $5.6 \rho_0$, stable hybrid stars exist with
a NM/NQM mixed phase in the center. For example, the maximum mass star
has a radius of $11.6\,$ km and the mixed phase is realized within $r=5.8$ km.
In the second case,
the transition to the NM/SQM mixed phase occurs at around $2.3 \rho_0$,
the pressure in the mixed phase is almost
constant, and around $4.6 \rho_0$  the neutral SQM phase is reached. 
Stable hybrid stars exist with central densities between $2.3 \rho_0$ and
$3.4 \rho_0$ with a very small region of NM/SQM mixed phase in the center and for central densities between $5.6 \rho_0$ and $13.4 \rho_0$ stable quark
stars may exist, which are composed of SQM in the central region. For example,
 the maximum mass star for the case $r_s=0.25$ has a radius of $8.2$ km and the SQM phase is realized within $r=6.0$ km. 
These results are qualitatively similar to the ones
reported in Ref.\cite{REF1A}. Comparison of two cases shown in Fig.3 
indicates that 
color superconductivity certainly can have a major effect on the EOS.  With
increasing values of $r_{s}$, both the density
and the pressure corresponding to the phase transition to QM are
significantly reduced, the EOS becomes considerably softer, and the 
star mass decreases for given central density.

It is interesting
to note that for all cases which we studied and where we have a
phase transition from neutral NM to neutral SQM (without admixing the NQM
phase),
the pressure changes only very little in the NM/SQM mixed phase. Then the
situation becomes similar
to the one where a naive Maxwell construction is applied after
imposing the neutrality condition for each phase separately.
Including the quark mass in the QM phase does not change our conclusions.
In the $r_s = 0.25$ case (which has the lowest transition density) the
mass causes only a slight decrease in the pressure ($\Delta P \approx 1
\rm{MeV} \rm{fm}^{-3} $) in the region of the mixed phase.  This is too
small an effect to be seen on the scale of the phase diagram and certainly
does not lead to any qualitative changes in our results.

We note that most of the recent calculations on SQM have been 
performed by using 
the 3-momentum cut-off scheme and in general the gap is found to be 
around 100 MeV in the intermediate density region \cite{GAP1}.
Our work differs in that we have used the proper time regularization scheme, 
which leads to larger values of the gap
(between 300 and 400 MeV in the relevant 
density region).  However, qualitatively the situation is 
similar to the cases of strong diquark coupling discussed in
Ref.\cite{REF1A,REF1B}.

Finally, we would like to come back to the important question of 
consistency between the values for $G_s$ used for the single nucleon
and for the pairing strength in SQM (see Table 1). In this work, we
fixed $G_s$ for the single nucleon by fitting the nucleon mass.
However, we have to note that there are further attractive contributions
to the nucleon mass, most importantly pion exchange 
\cite{Hecht:2002ej,Leinweber:2003dg,ISHII} and 
contributions of axial vector diquarks \cite{AX}, which would lead
to smaller values of $G_s$.
Indeed, using the expressions given in Ref.\cite{TW} for the
pion exchange contribution to the nucleon mass, we find that the
nucleon mass can be reproduced with $r_s=0.4(0.25)$, where the two
numbers refer to the case without and with further inclusion of
axial vector diquarks\footnote{
If axial vector diquarks are included $r_s$ is significantly decreased since there is alternative attraction through this channel. In this case $G_a$ (the axial coupling constant) is fixed using the Fadeev Equation for the $\Delta$ and setting $M_\Delta = 1232$ MeV.}. 
Thus the inclusion of pion exchange and axial vector diquarks for the nucleon mass will allow a common value of $G_s$ for the single nucleon and SQM.  
The details of this calculation will be discussed elsewhere \cite{FUT}.

\section{Summary}
\setcounter{equation}{0}

By applying a flavor SU(2) NJL model to both NM and QM phases, we have
studied the phase diagram for isospin asymmetric matter at finite density.
We emphasize that the model, and in particular the regularization
scheme which we used, describes the single
nucleon and the saturation of normal NM,
and therefore forms a basis to investigate the EOS at
higher densities.
We found that, as we vary the pairing strength in QM, several
scenarios are possible.  The charge neutral EOS may make
a transition to NQM or to SQM, via either one, two or three
globally charge neutral mixed phases.  These transitions begin at small
enough densities (2.3 - 3.4$\rho_{0}$) that the QM phase, or at least the
mixed phase, may occur inside neutron stars.

\vspace{0.5 cm}

{\sc Acknowledgments}

The authors wish to thank Dr.~David Blaschke for helpful
discussions.
This work was supported by the Australian Research Council and DOE
contract DE-AC05-84ER40150,
under which SURA operates Jefferson Lab, and by the Grant in Aid for
Scientific Research of the Japanese Ministry of Education,
Culture, Sports, Science and Technology, Project No. 16540267.

\end{document}